\documentclass{aastex}
\usepackage{emulateapj5, apjfonts}

\makeatletter
\newenvironment{tablehere}
  {\def\@captype{table}}
  {}

\makeatother

\newcommand\etal{{\it et~al.~}}
\newcommand\kms{ {\rm km\, s^{-1} } }
\newcommand\Ms{M_\odot}
\newcommand\Rs{R_\odot}

\begin{document}
\title{Radiation Hydrodynamical Evolution of Primordial H II Regions}
\author{Daniel Whalen}
\affil{Department of Physics, University of Illinois, Urbana-Champaign, Urbana,
IL  61801}
\email{dwhalen@cosmos.ucsd.edu}
\author{Tom Abel}
\affil{Pennsylvania State University, University Park, PA 16802}
\and
\author{Michael L. Norman}
\affil{University of California, San Diego, La Jolla, CA 92093}

\begin{abstract} We simulate the ionization environment of z $\sim$ 20
luminous objects formed within the framework of the current
CDM cosmology and compute their UV escape fraction. These
objects are likely single very massive stars that are copious UV
emitters.  We present analytical estimates as well as one--dimensional
radiation hydrodynamical calculations of the evolution of these first
HII regions in the universe. The initially D--type ionization front
evolves to become R--type within $\lesssim 10^5$\,yrs at a distance
$\sim1$\, pc. This ionization front then completely overruns the halo,
accelerating an expanding shell of gas outward to velocities in excess
of 30\,km\,s$^{-1}$, about ten times the escape velocity of the
confining dark matter halo. We find that the evolution of the HII
region depends only weakly on the assumed stellar ionizing
luminosities. Consequently, most of the gas surrounding the first
stars will leave the dark halo whether or not the stars produce
supernovae. If they form the first massive seed black holes these are
unlikely to accrete within a Hubble time after they formed until they
are incorporated into larger dark matter halos that contain more gas.
Because these I--fronts exit the halo on timescales much shorter than
the stars' main sequence lifetimes their host halos have UV escape
fractions of $\gtrsim 0.95$, fixing an important parameter for
theoretical studies of cosmological hydrogen reionization.

\end{abstract}


\section{Introduction}

In light of recent WMAP results on reionization \citep{ket03}
there has been renewed interest in early reionization scenarios
\citep{hh03,c03,sl03,bl03,soet03,cfw03,cfmr01}. The models rely heavily on  
UV production estimates for the
assumed stellar populations and the escape rates depending on the
global structure of the luminous proto--galaxies. The latter is
modeled by an escape fraction, f$_{esc}$, which is defined as the
fraction of ionizing photons produced by stars within a 
luminous object that exit the object. 

\subsection{Previous UV Escape Fraction Estimates}
Initial upper limits of only $0.02 - 0.15$ placed upon escape fractions for local 
starburst galaxies \citep{let95} raised concerns about the role early galaxies
could have had in reionization but these estimates were later revised 
upward by a factor of four \citep{hjd97}.  Lyman break galaxies may have escape 
fractions greater than 0.2 \citep{spa01}. Theoretical investigations 
usually derive very low escape fractions for even small high redshift 
galaxies in part because of the assumed scaling of the typical interstellar 
densities with $(1+z)^3$ \citep{wl00,rs00}. \citet{wl00} in particular do not
consider radiation-driven hydrodynamic motion in their static models that could 
open channels out the galaxy from which UV flux might escape.
A notable exception is the study by \citet{f02} which does model
galactic outflows with multidimensional hydrodynamical simulations and
finds an appreciably larger escape fraction of $\gtrsim 20\%$.  

Unfortunately, there is a clear limit to how far present upper bounds on
f$_{esc}$ can predict the escape of UV photons from Pop III stars because
the ionization environment of primordial stars residing in minihalos at
z $\sim$ 20 is very different from that observed in z $\lesssim$ 0.1 galaxies.
Zero-metallicity Pop III stars are not currently thought to have line-driven
stellar winds \citep{bhw01,kud00,vkl01} or exhibit instabilities that can drive 
significant mass loss on timescales smaller than the main sequence lifetimes
of these stars.  Consequently, the wind bubbles blown by OB associations in
galactic settings which can act to trap UV outflow from these stars \citep{dsf00}
are probably absent around Pop III stars so we do not consider them in our study.
Likewise, the primordial halos surrounding the first generation of stars were free
of dust that impedes escape of UV photons in modern galaxies.  UV escape fractions 
and their evolution in current models of early reionization have therefore remained
highly uncertain and been treated as a free parameter.   

Of similar importance in these 
models is the efficiency of UV photon production per baryon in stars. 
Interestingly, different stellar population synthesis models disagree on UV
production efficiency, regardless of IMF (see e.g. \citet{s02} or \citet{wa03} 
for a discussion).  Fortunately, high-resolution CDM hydrodynamical simulations now 
offer robust predictions of the masses and evolution of the first luminous
objects that can guide studies of early reionization.

\subsection{Geometry and Masses of Simulated First Luminous Objects}

Several different techniques have been applied to numerical
simulations of the first stars.  The initial studies using nested grid
Eulerian schemes \citep{a95,abet98} revealed that the first
cosmological objects condense cool 1000 $\Ms$ clouds in their
centers. At that time we argued that insufficent spatial resolution
and the exclusion of three body formation of molecular hydrogen
\citep{pss83,s83} left the fate of these large clouds uncertain.
Similarly medium-resolution smooth particle hydrodynamics simulations
\citep{bcl99} that followed the idealized collapse of uniformly
rotating spheres of masses of $2 \times 10^{6}\Ms$ also found cold
clouds of masses $\sim 10^3\Ms$ to form.  These rotating models first
create rotationally-supported disks that eventually fragment.  In
contrast, simulations that account for the full hierarchy of structure
in CDM models from realistic cosmological conditions produce no disks
\citep{a95,abet98,fc00,abn00,mba00,yet03}, independent of whether
smooth particle hydrodynamical or Eulerian adaptive mesh refinement (AMR)
techniques are employed.  Instead, these studies typically find the
first protostellar objects to be roughly spherical on scales from
$\sim 0.1$\,pc out to the virial radii ($\sim 100$\,pc), with masses
of approximately 100 $\Ms$ accreting within dark matter halos on the order of 10$^6$
$\Ms$ and forming by redshifts of 20 - 30. This absence
of global disks in the first galaxies and hence around the first
luminous objects is crucial to understanding their UV escape fraction
and the impact of the first supernova explosions. For a first study of
the formation and evolution of the first HII regions, however, it is
reasonable to employ only one dimensional radiation--hydrodynamical
models.

The highest resolution simulations to date with elements 
as small as $\sim 100$ $\Rs$ and $\sim 1/4$ M$_{\odot}$
\citep{abn02}, which also include three-body molecular hydrogen
formation, suggest that the primordal molecular cloud analogs do not
fragment. They instead form a single very massive star from the inside
out. The detailed processes limiting the accretion onto these first
protostars are not fully understood but \citet{abn02} suggested a
plausible mass range for the first stars to be between 30 and 300
$\Ms$.

Subsequent studies of \citet{op03} assert that the first stars might
be as large as 600 $\Ms$, at which point accretion times from
\citet{abn02} become equal to the lifetime of a primordial massive
star. However, as discussed in detail by \citet{hw02} such massive
stars would most likely bypass SN explosion to form black holes
directly without releasing the heavy elements necessary for the
formation of the next generation of lower-mass metal-enriched
stars. Omukai \& Palla further argued that these stars would never
achieve photospheric temperatures in excess of 6000 K to contribute UV
radiation to the early reionization of the universe.

The models of Omukai and Palla do not include full hydrodynamics and 
perhaps more
importantly only consider grey radiative transport using mean
opacities. Grey transport may fail to capture UV photons in the Lyman
Werner bands of H$_2$ which even in small numbers will destroy H$_2$
in the accreting envelope that cannot be reformed because of the
absence of dust and free electrons. This method may therefore fail to
properly cut off the H$_2$ cooling that permits accretion to continue,
leading to overestimates of protostellar mass. Considering the
three-dimensional nature of the accretion, it also seems unlikely that
the material would accumulate at the exact rate required to keep the
star from reaching the 100,000 K ZAMS temperatures predicted by theory
\citep{s02}.  However, see \citet{op03} and \citet{oi02} for a
markedly different view. We feel that only fully three-dimensional
radiation hydrodynamical simulations that follow radiative transfer
without resorting to grey opacities will be able to shed light on the
exact physics that presumably halt the accretion of the first
protostars.  Therefore, given that ab initio numerical simulations
yield the detailed structure of the first star-forming clouds and
good estimates for the range of expected luminosities of the first
stars are available, it is timely to discuss the properties of the
associated first HII regions. In this paper we 
demonstrate that the escape fraction from most microgalaxies
holding very massive primordial stars is of order unity. We make this
point with one dimensional radiation hydrodynamical simulations along
with purely analytical arguments. In light of the possible
overestimates of final primordial star masses in some current work we
only consider the 120 -- 500 $\Ms$ mass range of interest also to
future SN metal mixing studies.

\section{Reactive Flow and Radiative Transfer in ZEUS-MP}

\begin{figure*}
\epsscale{1.9}
\plottwo{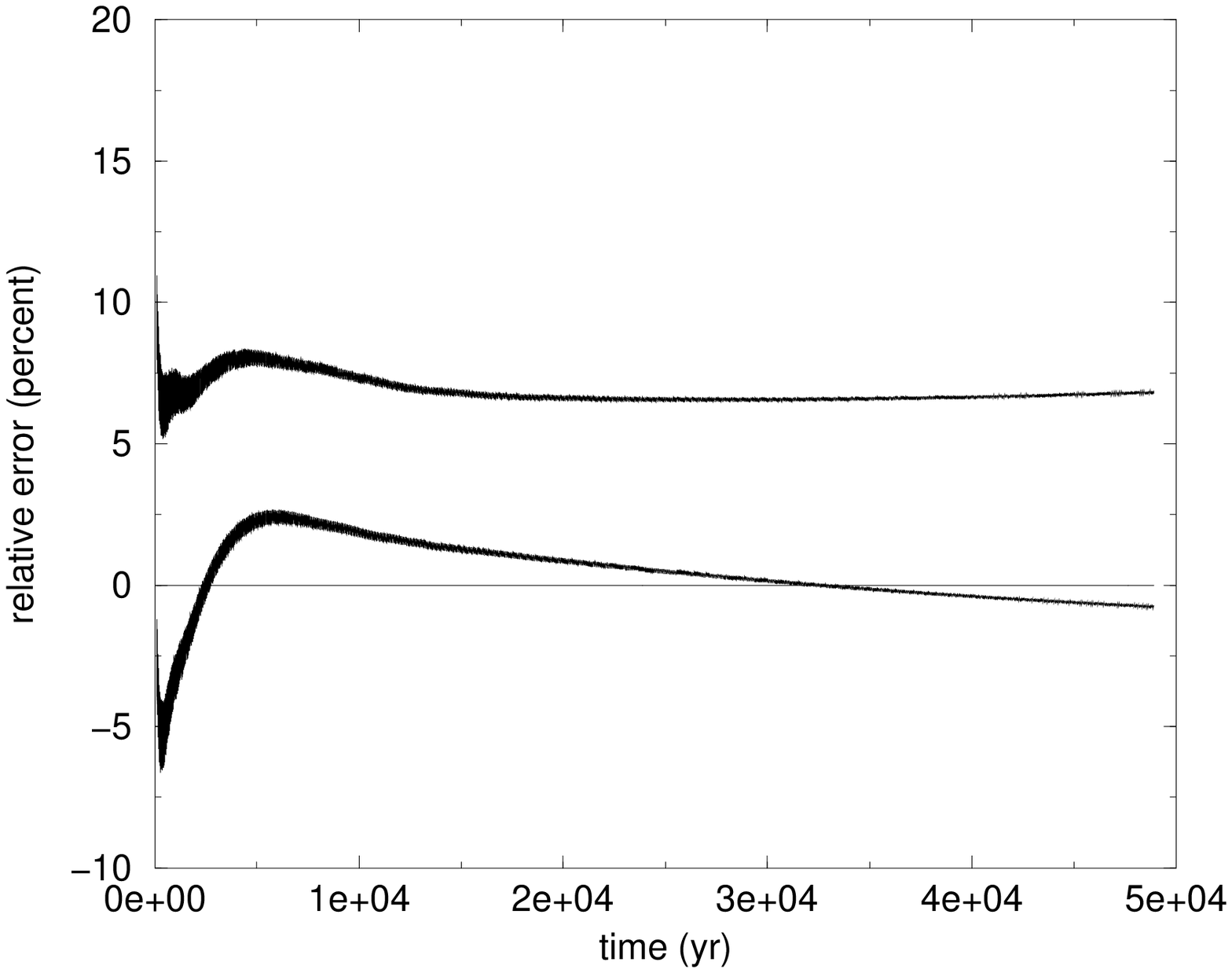}{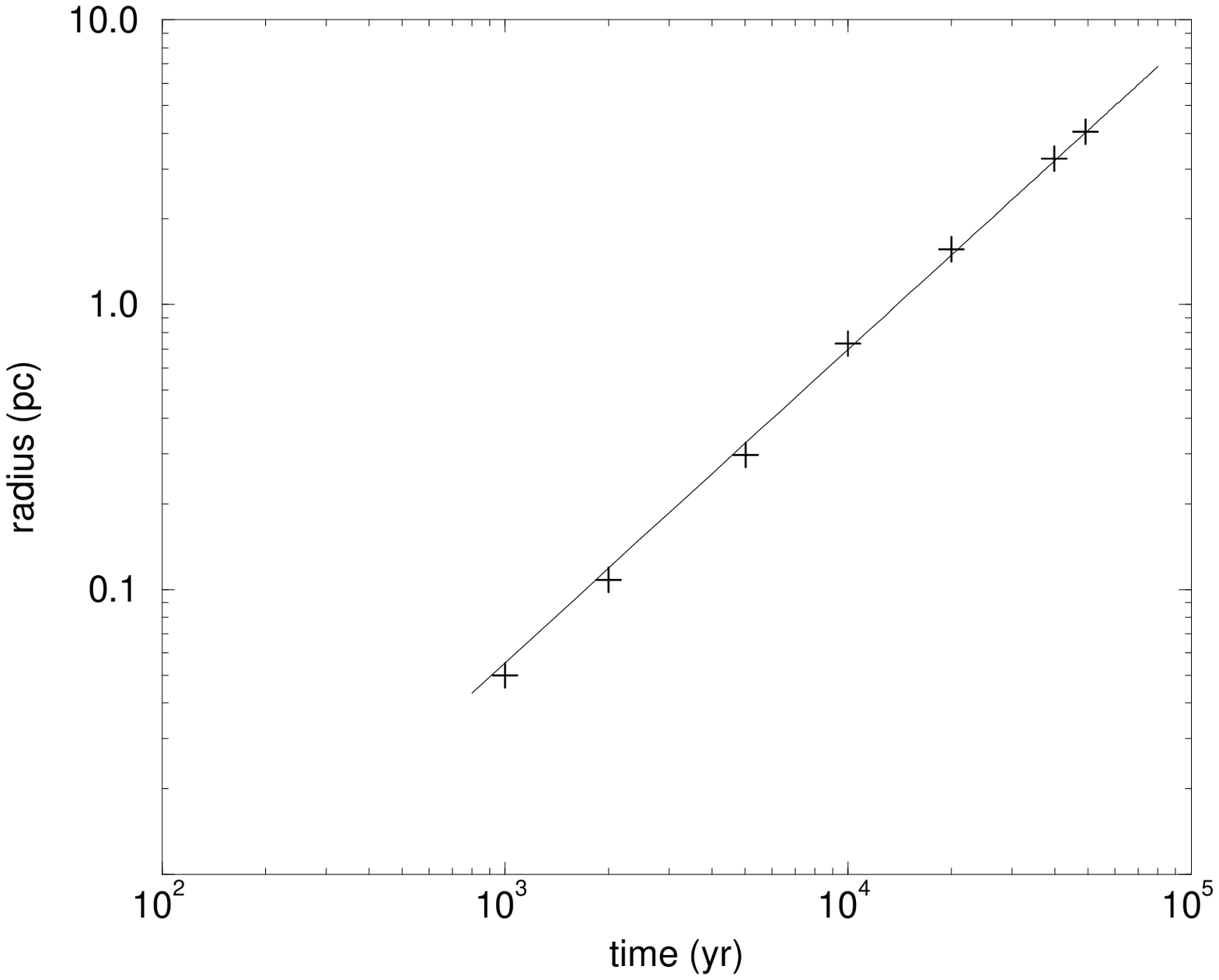}
\caption{ZEUS-MP propagation of I--fronts down r$^{-1}$ (left) and r$^{-3}$
(right) density gradients. The left panel displays the percent error in position
between the ZEUS-MP leading shock and I--front calculations and the \citet{ftb90} 
prediction. The upper and lower curves are the shock and front, respectively. The 
right panel compares ZEUS-MP core shock positions with theory (the I-front having 
already exited the halo).
\label{fig?}}
\end{figure*}
Collisional and ionization processes can ionize and recombine primordial
H and He into H$^{+}$, He$^{+}$, He$^{2+}$, H$^{-}$, H$^{+}_{2}$,
H$_{2}$, and e$^{-}$, which we added to ZEUS-MP by insertion of a
9--species reaction network and nine additional advection equations
\citep{abet97,anet97}.  For simplicity, the calculations we present here
assume the primordial gas is hydrogen only, which can be in neutral or
ionic form but not molecular.  We discuss the omission of helium later and
show that it has a minimal effect on our principal conclusions. The rate 
equation for change of neutral
hydrogen abundance (excluding the advection term) reduces to:\vspace{0.1in}
\begin{equation}
\displaystyle\frac{dn_H}{dt} = k_{rec}n_{H^+} n_{e} - k_{ph}n_{H} - k_{coll}n_
{H}n_{e}\vspace{0.1in}  
\end{equation}
We further modified ZEUS-MP to follow the radiative transfer of a
single point source of UV radiation at the center of a
spherical-coordinate geometry.  Photoionizations in any given cell are
due to direct photons from the central source entering the cell's
lower face as well as to diffuse photons entering through all its
faces. Recombinations within a zone occur to both the ground state and
to all the excited states. The on-the-spot approximation \citep{o89}
allows us to equate the diffuse ionizations in a cell to its
recombinations to the ground state, relieving us of the costly
radiative transfer from many lines of sight needed to compute the
diffuse radiation in the cell. The diffuse ionization and ground-state
recombination terms cancel each other in the network above so only a
single integration of the transfer equation along a radial ray from
the coordinate center is necessary to compute k$_{ph}$ in a zone.  

Recognizing that light-crossing times over most of the problem domain
are short in comparison to material evolution timescales enables us to
discard the time derivative in the equation of transfer that would
otherwise restrict the code to unnecessarily short timesteps. Rapid
ionizations very close to the central star will violate this
approximation and lead to superluminal I--front velocities, which we
prevent by computing fluxes only at distances smaller than
ct$_{problem}$ from the star (see \citet{anm99} for
discussion). In practice, the static approximation to the equation of
transfer usually becomes valid before the I--front approaches the
Str\"{o}mgren radius and the code computes Str\"{o}mgren radii and formation
times in excellent agreement with theory.

Hence, we solve the static equation of radiative transfer in flux
form\vspace{0.1in}
\[ \begin{tabular}{cc} 
$\nabla \cdot \bf{F} = - \chi\, {\bf{F}}_r$, & $\chi = \displaystyle\frac
{1}{n \sigma}$
\end{tabular} \vspace{0.1in}  \]
to produce the central flux piercing the bottom of each cell\vspace{0.1in}
\begin{equation}
F_{i} = \left(\displaystyle\frac{r_{i-1}}{r_i}\right)^{2} F_{i-1}\, e^{{-\chi_i 
\left(r_{i}-r_{i-1}\right)}}\vspace{0.1in}
\end{equation}
which yields the photoionization rate coefficient k$_{ph}$ with units 
$[1/{\rm s}]$ for the i$^{th}$ radial zone:\vspace{0.1in} 
\begin{equation}
k_{ph}=\,  \displaystyle\frac{F_{i}\, \left( 1 - e^{{-\chi_i 
\left(r_{i+1}-r_{i}\right)}}\right)\,A_{inner}}{{h\nu}\,{n_{H}\,V_{cell}}}
\vspace{0.1in}
\end{equation}
where V$_{cell}$ and A$_{inner}$ are the volume and area of the lower
face of the cell.  This numerical scheme ensures conservation of
photons in all problem zones, a highly-desirable property that ensures
I--fronts propagate at speeds that are independent of problem
resolution.  We consider only monochromatic radiative transfer,
wherein a fixed amount of heat $\epsilon_{\Gamma}$ is deposited by the
radiation per photoionization. Though not necessary for 1-D studies,
this restriction greatly reduces computational cost and will allow full
three-dimensional simulations (Whalen \etal in preparation). 

Photoionization, collisional ionization and recombination processes can occur
over highly disparate timescales that are all in general much shorter
than the hydrodynamic response of the gas.  We finite difference the
reaction network in a semi--implicit scheme which employs densities
at the advanced time but rate coefficients evaluated at the current
time. This algorithm bypasses costly matrix iteration that would
become prohibitive in 3--D while retaining sufficient accuracy to
follow I--front propagation.  The code subcycles the reaction network
and isochoric energy equation by a tenth of the photoionization
timestep\vspace{0.1in}
\begin{equation}
t_{chem} = \displaystyle\frac{n_{e}}{{\dot{n}}_{e}}\vspace{0.1in}
\end{equation}
until it has covered a tenth of the heating/cooling timestep\vspace{0.1in}
\begin{equation}
t_{heat/cool} = \displaystyle\frac{e_{gas}}{{\dot{e}}_{heat/cool}}\vspace{0.1in}
\end{equation}
at which point it updates the hydro equations. 

The absence of metals in the first-generation stars and their primordial
envelopes permits the use of simple cooling rates in the ischoric energy
equation.  Only photoionization, recombination, electron collisional 
ionization and excitation occur in the atomic hydrogen we consider in 
our calculations, yielding the energy equation:
\begin{equation}
{\dot{e}}_{gas} = k_{ph}\epsilon_{\Gamma}n_{H} - \Lambda_{rec}
n_{H^+} n_{e} - \Lambda_{coll\:ioniz}n_{H}n_{e} - \Lambda_{coll \:exc}n_{H}n_{e}\vspace{0.1in}  
\end{equation}
where k$_{ph}$ is the photoionization rate described above, $\epsilon_{\Gamma}$ is
the fixed energy per ionization deposited into the gas (set to 2.4 eV for reasons
explained in Section 5), and $\Lambda_{rec}$, $\Lambda_{coll\: ioniz}$, and 
$\Lambda_{coll\: exc}$ are the recombinational and collisional cooling rates taken
from \citet{anet97}). These four processes act in concert with hydrodynamics (such as
adiabatic expansion or shock heating) to set the temperature of the gas everywhere 
in the simulations.

\section{Code Tests}

We adopted the benchmarks that Franco \etal \citep{ftb90,tet86} employed in 
their theoretical studies of I--front propagation, in which a source of 
ionizing photons was centered in a radial density profile with a flat core
and an r$^{-\omega}$ dropoff:\vspace{0.1in}
\[ n_{H}(r) = \left\{ \begin{array}{ll}
			   n_{c}                 & \mbox{if $r \leq r_{c}$} \\
			   n_{c}(r/r_{c})^{-\omega}   & \mbox{if $r \geq r_{c}$}
                          \end{array}
                  \right.\vspace{0.1in} \]
We applied the same initial density profiles, central source strength,
and cooling physics in our tests as did Franco, \etal Their studies
of I--front propagation down these radial density gradients indicate
there is a critical exponent $\omega = $ 1.5 below which the front
executes the classic approach to a Str\"{o}mgren sphere, reverts from
R-type to D-type, and subsequently drives a dense shocked shell before
it thereafter down the density gradient \citep{ftb90}. The front
remains D-type throughout its evolution and accumulates mass in its
shell for as long as it expands. Fig 1 compares ZEUS-MP's propagation
of a D-type I--front down an r$^{-1}$ density gradient to the
predicted analytical result.  The code results agree with theory to
within 7\% at early times and 3\% at later times. We expect slightly
greater disagreement at early times because no current theory
addresses the hydrodynamic details of the R- to D-transition but
becomes increasingly accurate as the I--front evolves. The shock front
can be seen to precede the I--front with the analytical result (which
assumes the shock and front positions essentially coincide) falling
between them.
\begin{figure*}
\epsscale{1.0}
\plotone{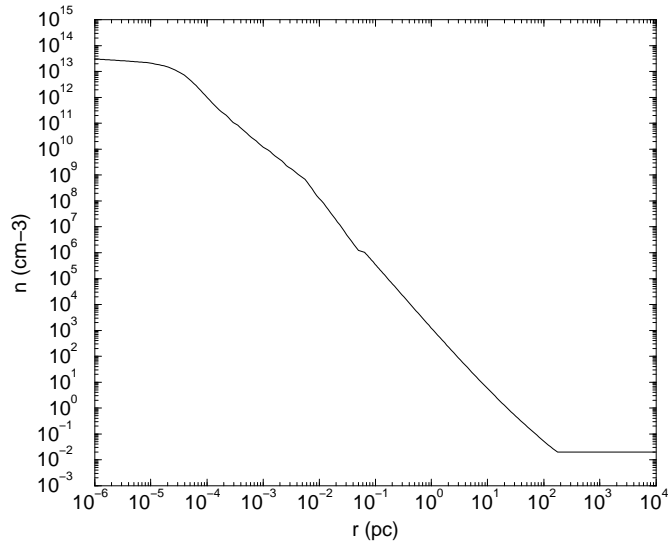}
\caption{Primordial cloud number density from \citet{abn02}. \label{fig?}} 
\end{figure*}

Ionization fronts descending density gradients steeper than the critical $\omega 
= $ 1.5 rapidly revert from D-type back to R-type (in some cases never having 
switched from R-type in the first place) and quickly overrun the cloud. Now 
completely ionized and at much higher pressures, the entire cloud inflates outward
at the sound speed of the ionized material, typically on the order of 10 km s$^{-1}$.
In this scenario strong core density gradients in the now mostly-isothermal cloud 
become strong pressure gradients which drive a shock that pistons the already-moving 
cloud material outward in a champagne flow. The shocked core expands at nearly 
constant velocities down 1.5 $<$ $\omega$ $<$ 3 gradients but strongly accelerates 
out through $\omega$ $>$ 3 gradients.  Runs conducted with ZEUS-MP for 1.5 $<$ 
$\omega$ $<$ 5 demonstrate this behavior. Fig 1 compares ZEUS-MP's expansion of the
ionized core to the core expansion rates of the \citet{ftb90} theory, demonstrating 
agreement to within 10\% at early times and 2\% at later times.

\section{Simulation Set Up}

We show in Fig 2 the fit we computed to the spherically-averaged gas number 
density surrounding the primordial protostar of the very high dynamical range 
AMR calculation performed by \citet{abn02}. In that calculation the protostellar 
core accreted within a 7 $\times$ 10$^5$ M$_\odot$ dark matter halo by a redshift of
18.2.  At r $\sim$ 175 pc we smoothly joined this fit to a constant baryon number 
density of 0.02 cm$^{-1}$, which is roughly 10 times the cosmic mean at this redshift, 
to approximate the higher circum-halo density found in \citet{abn02}.  In Table 2 we 
list final I-front radii assuming a circum-halo density equal to the cosmic mean.  
The spherical averaging of the baryonic density removes clumping and other features 
whose effects will be studied when we extend these simulations to 3-D in a forthcoming 
paper. 
%
%

The density was assigned a uniform temperature of 300 K (typical of the
virialized baryonic matter cooled by molecular hydrogen in the \citet{abn02}
simulations) and made
static by applying the gravitational potential necessary to hold it in
hydrostatic equilibrium (this potential would be that exerted by the
dark matter halo were it explicitly present in our simulation). We
evolved this model without a central ionizing source to make certain
the gas is in sufficiently stable equilibrium and found peculiar
velocities to be less than 0.1 $\kms$ after the three million years
typical for lifetimes of very massive primordial stars.  The
gravitational potential was held constant throughout the evolution of
the problem because the exit of the gas from the halo should have
minimal impact on the more massive equilibrium dark matter halo left
behind. For the results presented below we used 142 logarithmically-spaced 
radial zones. We placed a series of constant ionizing luminosities shown 
in Table 1 corresponding to a range of stellar masses and main-sequence 
lifetimes \citep{s02} at the center of the distribution (the 120 M$_\odot$
star was what emerged from the \citet{abn02} calculations).
 
%
\begin{tablehere}
\begin{center}
\begin{tabular}{c}
Table 1: Ionizing Photon Rates and Main Sequence Lifetimes \\
                                            \\  \end{tabular}
\vspace{-0.1in}
\begin{tabular}{lcc}
\tableline\tableline
\footnotesize{M$_\odot$} & \footnotesize{main sequence lifetime (yrs)} & 
\footnotesize{${\dot{n}}_{ph}$}  \\
\tableline
\footnotesize 500   &   \footnotesize 1.899e06    &  \footnotesize 6.802e50  \\
\footnotesize 400   &   \footnotesize 1.974e06    &  \footnotesize 5.247e50  \\
\footnotesize 300   &   \footnotesize 2.047e06    &  \footnotesize 3.754e50  \\
\footnotesize 200   &   \footnotesize 2.204e06    &  \footnotesize 2.624e50  \\
\footnotesize 120   &   \footnotesize 2.521e06    &  \footnotesize 1.391e50  \\
\tableline
\end{tabular}
\end{center}
\vspace{0.1in}
\end{tablehere}
\noindent These constant hydrogen ionization luminosities were obtained by averaging 
the stars' time-varying outputs over their main sequence lifetimes, assuming no
mass loss \citep{s02}. Time-dependent luminosities are easily incorporated into the 
code and will be included in later 3D studies where their effects may be more 
pronounced.

\begin{figure*}
\epsscale{1.4}
\plotone{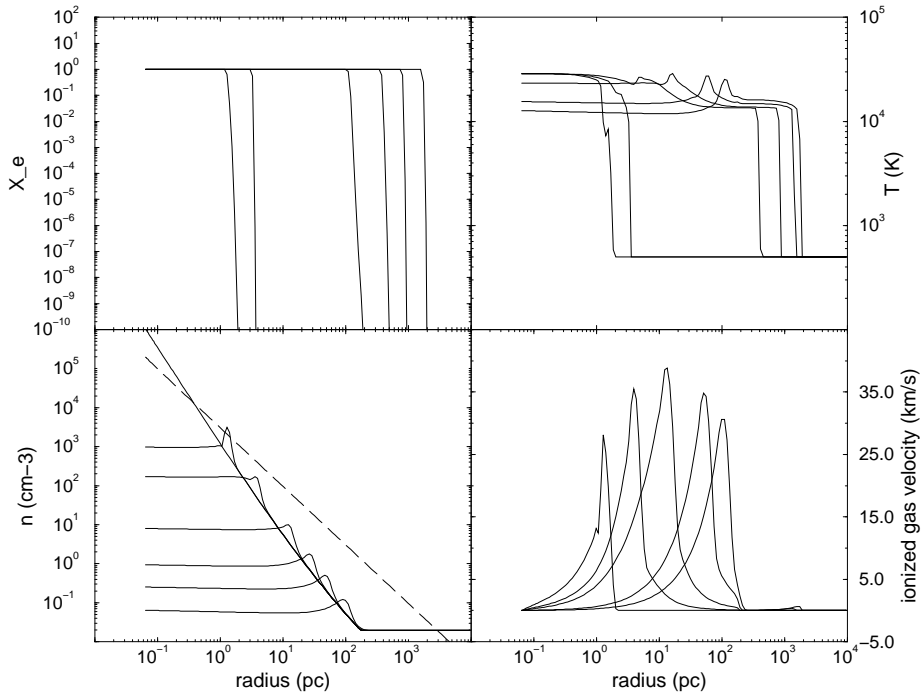}\vspace{0.15in}
\caption{Expansion of an HII region from a 200 M$_\odot$ star. The ionization 
fraction profiles are shown for t = 63 kyr, 82 kyr, 95 kyr, 126.9 kyr, 317 kyr, and 
2.2 Myr. The temperature distributions are output for t = 63 kyr, 82 kyr, 126.9 kyr, 
317 kyr, 1.11 Myr, and 2.2 Myr, and the densities are shown for t = 0 yr, 63 kyr, 
126.9 kyr, 317 kyr, 634 kyr, 1.11 Myr, and 2.2 Myr. The velocity profiles shown are 
for t = 63 kyr, 126.9 kyr, 317 kyr, 1.11 Myr, and 2.2 Myr. The dashed line in the 
density panel is the Str\"{o}mgren density (the density required to form a Str\"{o}mgren 
sphere and therefore initially bind the I--front) at a given radius. \label{fig?}}
\end{figure*}

\section{Results and Discussion}

\subsection{A Standard Case}

In non--rotating stellar evolution models stars above 50 M$_{\odot}$
and below 140 M$_{\odot}$ as well as above 260 M$_{\odot}$ likely
avoid supernova explosions and collapse directly into black holes
\citep{hw02}. We examined first the HII region formed by a 200 
M$_{\odot}$ star to understand the environment into which a primordial 
pair--instability supernova would detonate to produce metal enrichment.
Fig 3 displays the ionization, temperature, density, and velocity
profiles of the HII region expanding outward from this star for
the output times listed. Initially the D--type
ionization front is trapped as expected \citep{ftb90} and a shock
forms ahead of the front. The first two temperature profiles capture
the I--front just as it reverts from D-type back to R-type. At 82 kyr
the 25000 K front has just overtaken the 7000 K shock that had been
leading at 63 kyr.  The I--front then abruptly descends the density
gradient from 3 pc to 100 pc over the next 15 kyr as seen in the third
ionization profile, setting all of the ionized gas into outward motion. 
The strong density jump just before the front at 63 kyr is gone at 126 
kyr as the front advances well ahead of the shock leaving essentially 
undisturbed (though much hotter) gas in its wake.

One can apply a simple analytical argument to obtain a rough estimate of where the 
front reverts to R-type by computing the Str\"{o}mgren density at a given radius assuming
no hydrodynamic motion:
\begin{equation}
n_S = \left(\displaystyle\frac{3 F_{\ast}}{4 \pi {\alpha}_{B}}\right)^{\frac{3}
{2}} 
r^{-{\frac{3}{2}}}, \vspace{0.1in}
\end{equation}
where F$_{\ast}$ is the number of ionizing photons per second from the
star.  This is the density required to stall the early front to a
Str\"{o}mgren radius before hydrodynamic motions free it to advance forward.
n$_{S}$ appears as the dashed line on the density panel in Fig 3.  The
initial density profile can temporarily trap the I--front wherever it
is greater than n$_{S}$ but not when it dips below it, so one would
expect the I--front to revert to R-type beyond that radius. In reality 
this transition is delayed when the front encounters and must also
ionize the overdense shock. The delay is evident when comparing the 
ionization and density profiles because the t = 0 density drops below 
n$_{S}$ at 0.4 pc but the front does not overtake the shock until 3 pc. 

Once again R-type, the front rapidly inflates outward on timescales 
much shorter than the hydrodynamic response of the gas. The gas motion 
that follows this abrupt ionization
becomes apparent in the later temperature, density and velocity
profiles in Fig 3. Just as in the Franco benchmarks, strong density
gradients in the now ionized and nearly isothermal gas (as seen in the
temperature plots) become strong pressure gradients beyond the edge of
the density's flat central core which drive a shock out into the
ionized gas. The shock weakly accelerates but achieves velocities in
excess of 35 km s$^{-1}$ in comparison to the 12 - 15 km s$^{-1}$
sound speed of the 15000 K gas and the 2 - 3 km s$^{-1}$ escape
velocity from the halo. The small velocity bumps beyond 1 kpc in the
final two velocity profiles are due to the outer edge of the front
slowing down to a new Str\"{o}mgren radius as it traverses the constant
interhalo mean density. The velocities growing there will steepen 
into a shock behind the front that would eventually overtake it if the
central star continued to shine.

The temperature peaks at later times that coincide with the velocity 
peaks are from shock heating. The postshock gas temperature gradually
decreases with time from adiabatic expansion, recombinational
cooling, and electron-ionization, and -excitation cooling processes in
the primordial hydrogen. The preshock gas is hotter than the postshock
gas since it has not had as much time to cool and will be heated by
the shock's passage. 
\begin{figure*}
\epsscale{1.7}
\plottwo{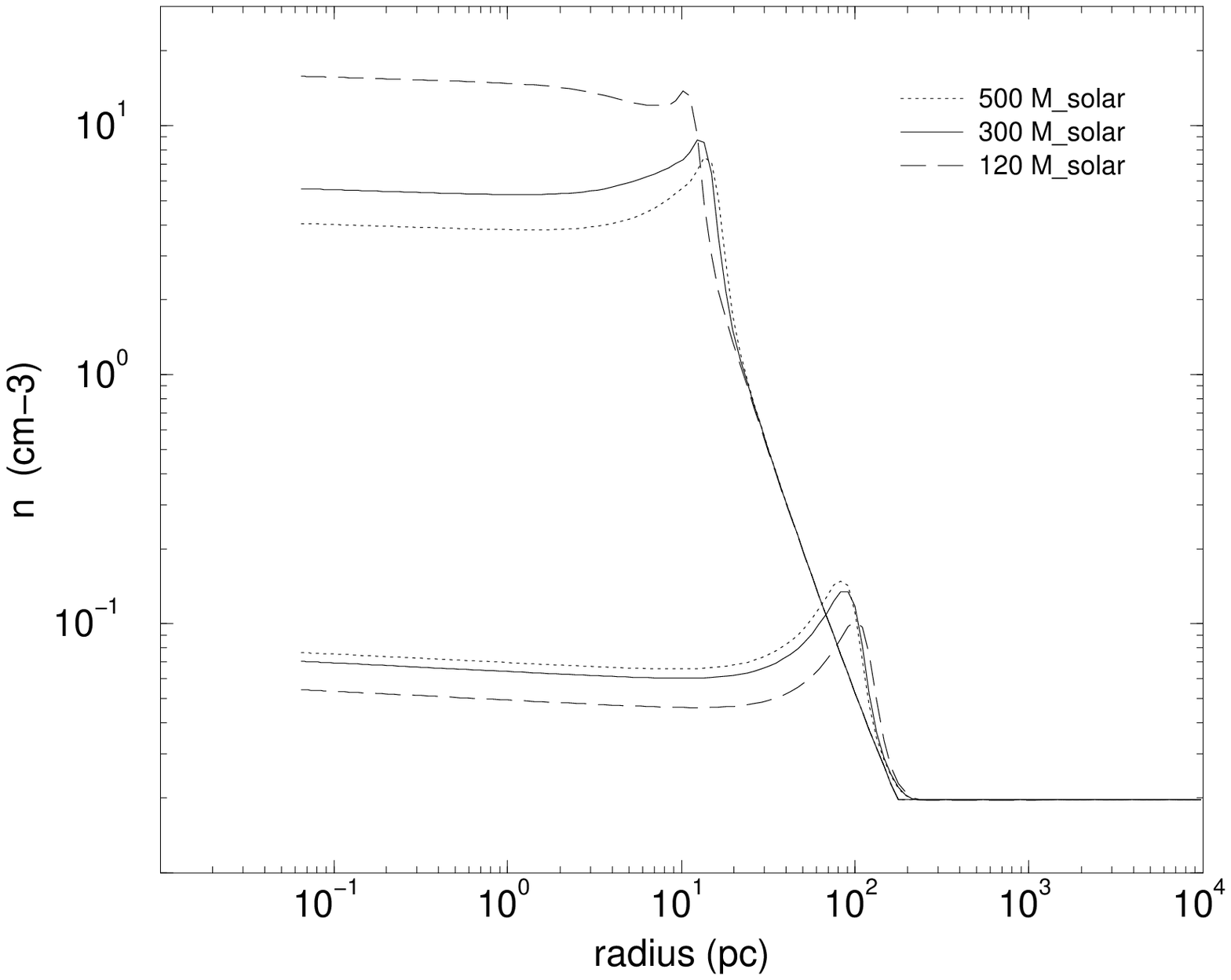}{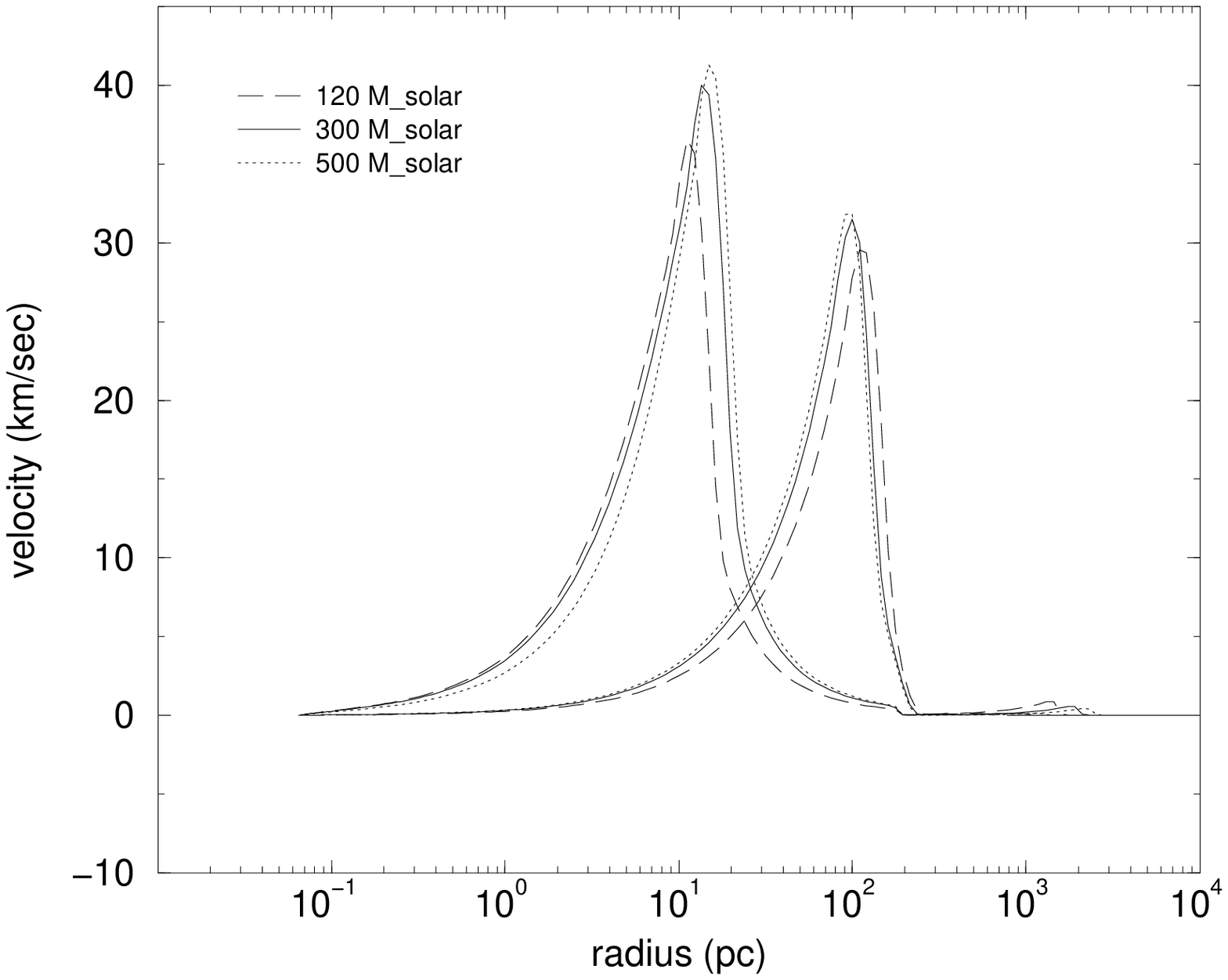}
\caption{Left panel: HII region density profiles for 120 M$_\odot$,
300 M$_\odot$, and 500 M$_\odot$ central stars. The densities shown
are for 317.1 kyr and then again at the main sequence cutoff time of
each star: 2.5, 2.0 and 1.9 Myr, respectively.  Right panel: Velocity
profiles for 120 M$_\odot$, 300 M$_\odot$, and 500 M$_\odot$ central
star HII regions. The velocities shown are for 317.1 kyr and then
again at the main sequence cutoff time of each star: 2.5, 2.0 and 1.9
Myr. \label{fig?}}
\end{figure*}

The time-sequenced density profiles clearly show the flow of
photoionized gas out of the halo, with the flat postshock densities
gradually draining away to twice the interhalo mean. The shock and
density peaks reach the 100 pc virial radius of the halo by the star's
main sequence lifetime. Nearly half the gas originally interior to
100 pc has flowed out by this time as seen in Fig 5, which plots the
total gas mass enclosed within a given radius for the three output times
listed. The uppermost curve is the original undisturbed halo
gas. The difference between this curve and the later ones at any
radius is a measure of how much material has been driven from that
radius. At 2.2 Myr 6000 M$_{\odot}$ of the initial 15000 M$_{\odot}$
present has crossed over 100 pc. The ionized flow is efficient at
expelling gas from the interior of the halo, leaving its innermost 50
pc at only twice the intergalactic mean density. Because the shock sweeps the
bulk of the expelled material outward at velocities well in excess of
the 2 - 3 km s$^{-1}$ escape velocity from the halo, its return
to the halo on merger timescales is unlikely.

\subsection{Varying the Ionizing Luminosity}

We computed four additional cases having the same initial density
profile but with the different luminosities in Table 1. The HII
regions in each case follow the same evolutionary stages as in the 200
M$_{\odot}$ run. HII densities and velocities for 120 M$_{\odot}$, 300 
M$_{\odot}$, and 500 M$_{\odot}$ central stars are shown in Fig 4, first 
at 0.6 Myr and then at the main sequence cutoff times 2.5, 2.0 and 1.9 
Myr of each, respectively.

It is immediately apparent from both plots that the ionized flow's characteristics
are largely independent of central photon rates. The peak shock speeds differ by at 
most 5 km s$^{-1}$ even though the luminosities vary by a factor of 5. The relatively
minor variation in shock velocities leads to the small spread in shock positions
seen at the two times in the density plot. The R-type front leaves the halo ionized 
on timescales short in comparison to the subsequent flow times so the initial ionized
density left in its wake is not very different from the original profile, the same 
used in all the luminosities we considered. Because it is the density and pressure 
gradients left by the front (and not the front itself or the luminosity that drove 
it) that govern the hydrodynamics of the resultant photoevaporation, it is reasonable
to expect different luminosities to deliver the same ionized flows. I--front dynamics 
in the core are responsible for the spread in shock speeds. Fronts 
driven by greater luminosities leave slightly steeper ionized densities behind at the 
core edge. The accompanying steeper pressure falloff accelerates the gas to the slightly
greater speeds seen in the higher luminosity runs.

At 0.6 Myr the postshock densities are lowest for the brightest star
as one might expect because its somewhat faster core shock pushes
matter from the halo more quickly. In contrast, the dimmest star
leaves the smallest densities in the halo at its main sequence
lifetime, but only because the same photoevaporative outflow present in
all three cases has been acting for the longest time. Our simulations show
that the ionized outflow from the more massive stars persists well
after their shutoff. In Fig 6 we plot the density profiles left by the
three stars at 2.5 Myr, shutting down the central source after the
star's MSL but permitting recombinational, electron collisional, and
hydrodynamical processes to continue. 
%
%
%
\begin{figure*}
\epsscale{1.0}
\plotone{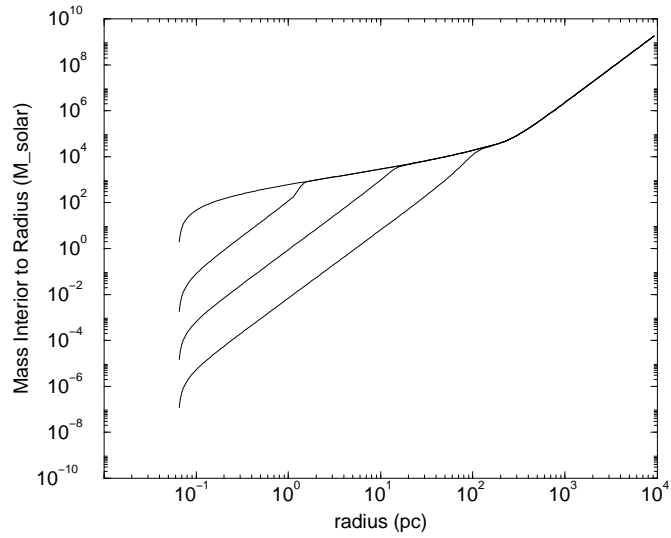}
\caption{Mass interior to a given radius for t = 63 kyr, 317 kyr, and 
2.2 Myr. The uppermost line is for the initial undisturbed density. \label{fig?}} 
\end{figure*}
If allowed to act in the absence of a supernova, post-MSL outflow from the more massive 
stars would evacuate the halo more completely than the longer-lived dimmer stars, as 
also evident in Fig 5.

We reran these calculations with the same halo but with an intercluster baryon density of
0.00135 corresponding to a redshift of 18.2 for a standard cosmology with 
$\Omega_B\,h^2 = $ 0.0224 consistent with the recent WMAP results \citep{spet} 
and the primordial deuterium abundance measurements by \citet{bt98} to compute the 
final volumes ionized by the five luminosities.  The final I--front position for each 
is listed in Table 2. As expected, the brightest stars ionize the greatest volumes.
\begin{tablehere}
\begin{center}
\begin{tabular}{c}
              Table 2: Final I--front Radii               \\
\vspace{-0.1in}
\end{tabular}
\end{center}
\end{tablehere}
\begin{tablehere}
\begin{center}
\begin{tabular}{cc}
\tableline\tableline
\footnotesize{M$_\odot$} & \footnotesize{final I--front position (pc)}   \\
\tableline
\footnotesize 500   &   \footnotesize 5363    \\
\footnotesize 400   &   \footnotesize 5080    \\
\footnotesize 300   &   \footnotesize 4558    \\
\footnotesize 200   &   \footnotesize 3874    \\
\footnotesize 120   &   \footnotesize 3120    \\
\tableline
\end{tabular}
\end{center}
\vspace{0.1in}
\end{tablehere}
%
%

%
\begin{figure*}
\epsscale{2.0}
\plottwo{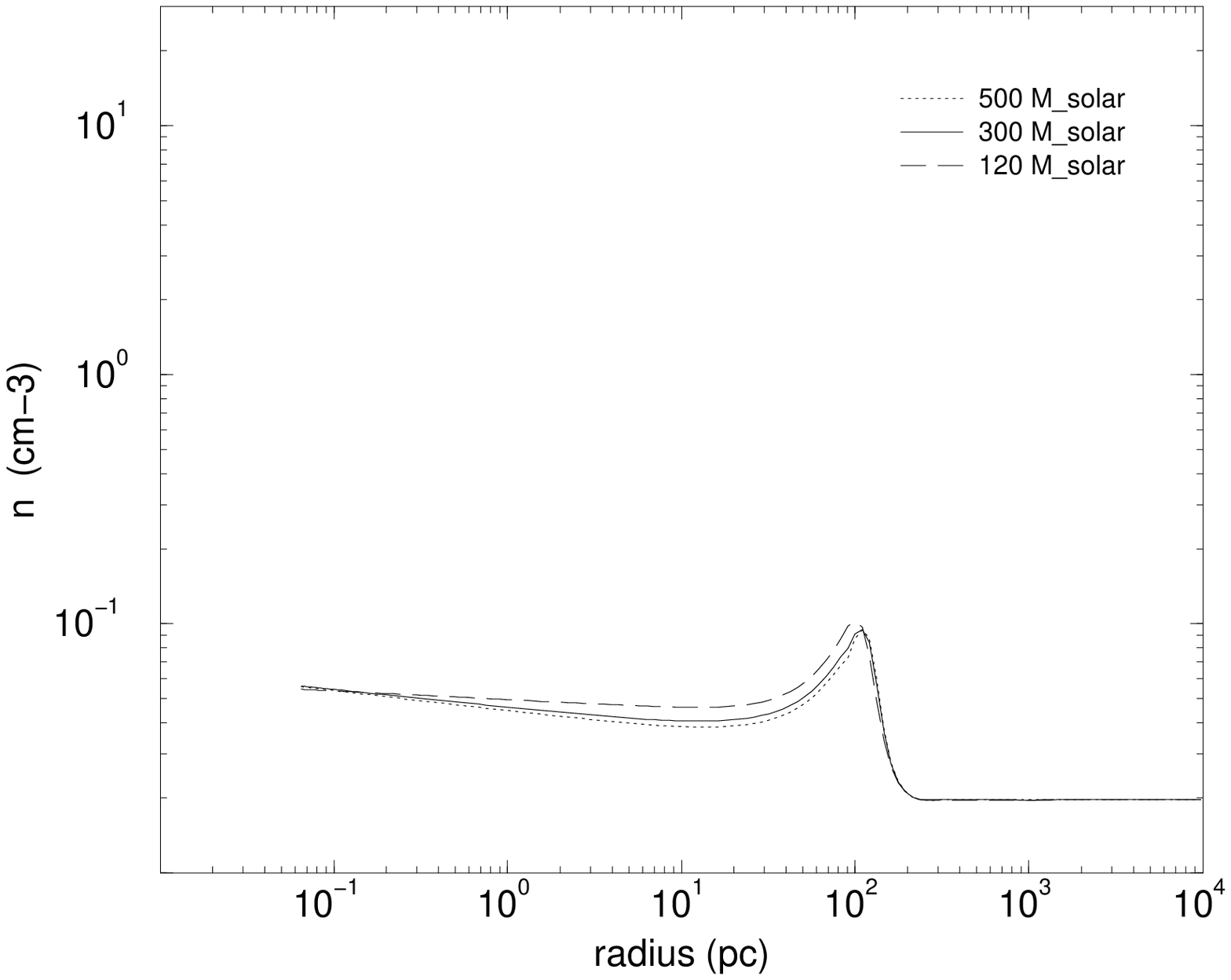}{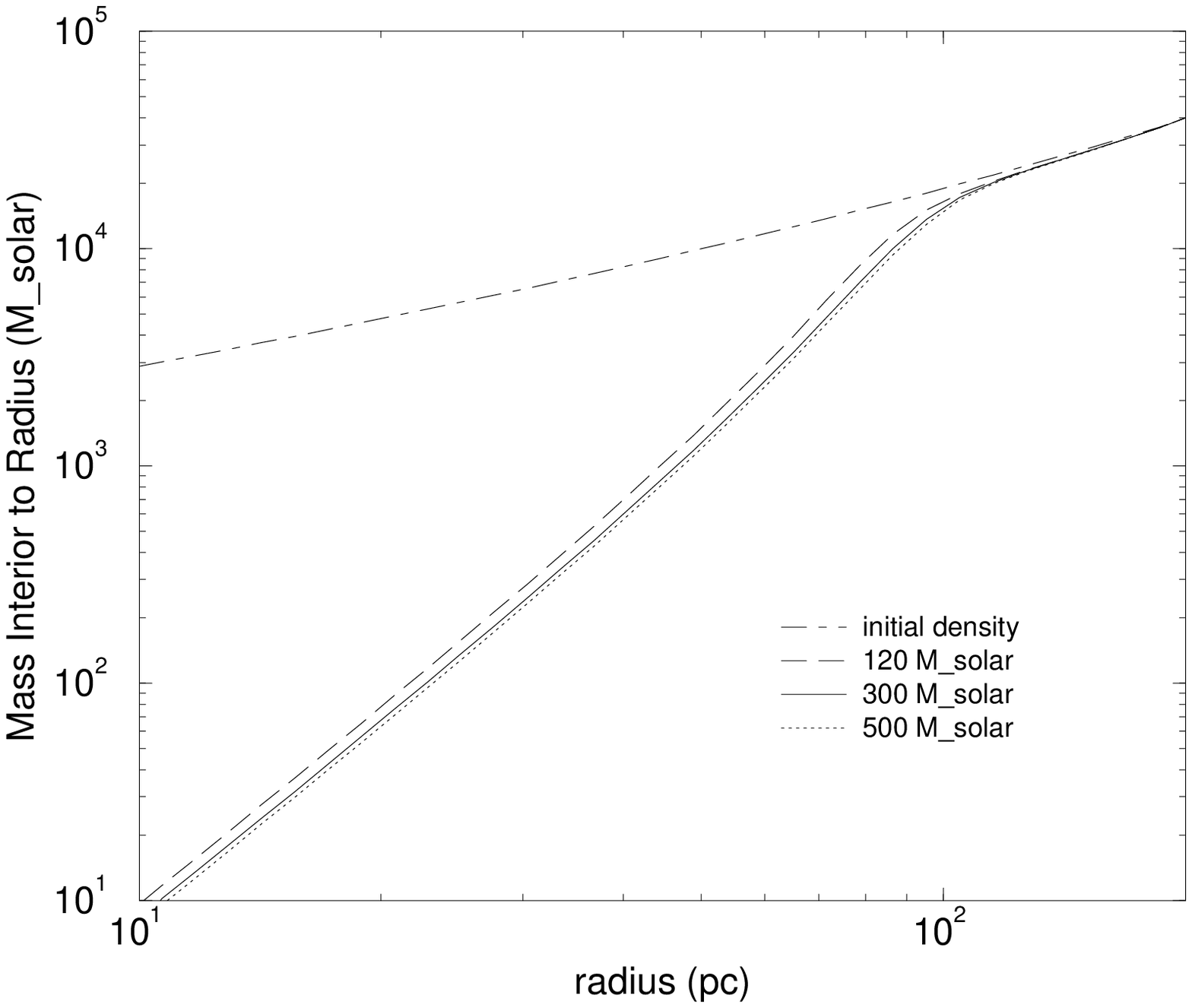}
\caption{Left panel: Density distributions of 120 M$_\odot$, 300 M$_\odot$, and 500 
M$_\odot$ star HII regions permitted to flow out to 2.5 Myr. Right panel: Gas mass 
interior to a given radius again, but at the 2.5 Myr main sequence lifetime of a 120 
M$_\odot$ star. \label{fig?}}
\end{figure*}
\begin{figure*}
\epsscale{1.4}
\plotone{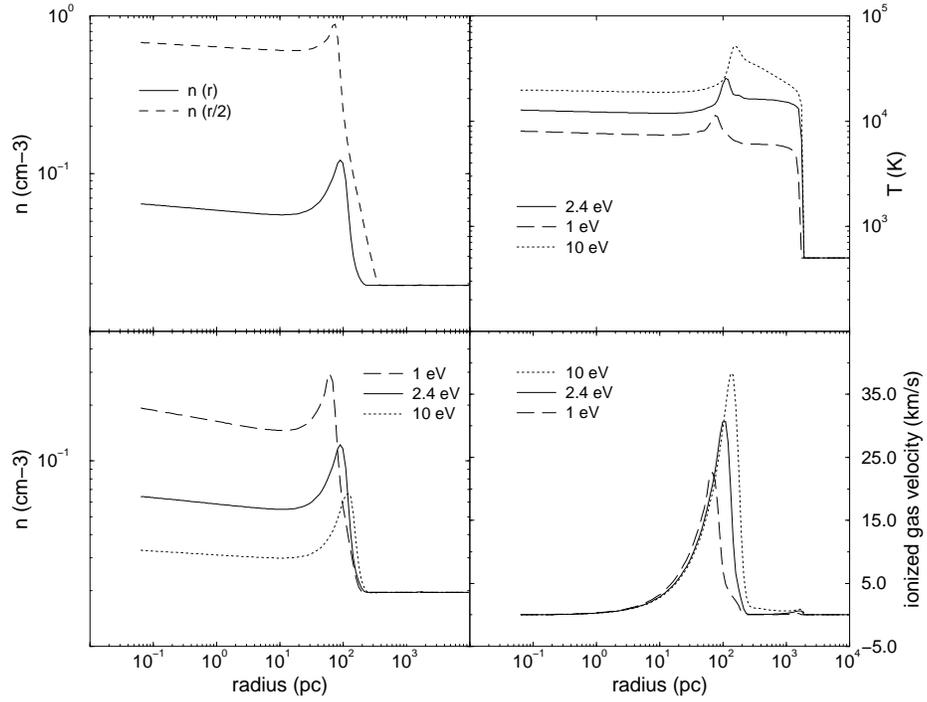}
\caption{Upper left panel: HII region density profiles for a 200 M$_\odot$
central star residing in the original \citet{abn02} halo density profile n(r) 
and within a more massive halo n(r/2). The densities shown
are for 2.2 Myr.  Upper right panel: temperature profiles for 1 eV, 2.4 eV, and 
10 eV energy depositions into the gas, taken at 2.2 Myr. Lower panels: density
(left) and velocity (right) plots for 1 eV, 2.4 eV, and 10 eV at 2.2 Myr.
\label{fig?}}
\end{figure*}
\begin{figure*}
\epsscale{1.0}
\plotone{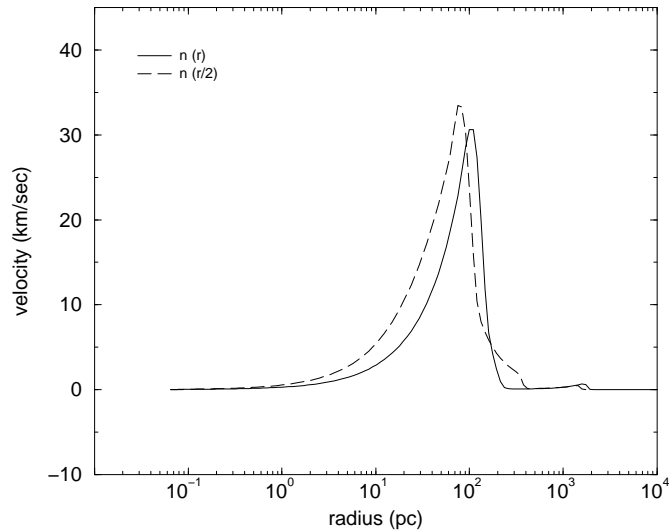}
\caption{Ionized core shocks in the original \citet{abn02} halo density profile 
n(r) and in a more massive halo n(r/2) \label{fig?}}
\end{figure*}
%
%

%
%

%
%

%
%

\subsection{He Ionization}

Our use of monochromatic radiative transfer underestimates the UV 
photoheating of the halo envelope by not accounting for the greater
average energies imparted to electrons by the photons well above 
threshold emitted by the Pop III VMS. If only hydrogen 
were present this would be a relatively minor effect because of the 
${\nu}^{-3}$ falloff of the ionization cross section. Neglecting 
primordial He ionization is more serious because it is heated to much
higher temperatures than hydrogen by the harder photons in the stellar
spectrum, though there are less of these photons as seen in Table 3.

\citet{ah99} have shown that one can approximate the heat input of He 
ionization by increasing the fixed amount of heat 
$\epsilon_{\Gamma}$ deposited by the radiation per photoionization. They
found the amount by which this heat should be increased is sensitive to 
the radiation environment but in general is between 0.5 E$_{ioniz}$ and 
E$_{ioniz}$. Our calculations to this point all applied 2.4 eV of heat per ionization
to the gas. Temperature, density, and velocity profiles for a 200 
M$_{\odot}$ star for energy depositions of 1 eV, 2.4 eV, and 10 eV per 
ionization are shown in Fig 7. Helium ionization raises the temperature
to between 20,000 K and 50,000 K in the front and to 20,000 K behind the
front. The accompanying increase in sound speed results in the higher
shock speeds and lower densities seen in the 10 eV case, demonstrating
that the main contribution of He ionization is to accelerate the 
photoevaporative flow of gas from the halo. Our results for 2.4 eV   
therefore represent a lower limit to this flow.

\begin{tablehere}
\begin{center}
\begin{tabular}{c}
Table 3: H, He, and He$^{+}$ Ionizing Luminosities \citep{s02}  \\
                                            \\  \end{tabular}
\vspace{-0.1in}
\begin{tabular}{lccc}
\tableline\tableline
\footnotesize{M$_\odot$} & \footnotesize{Q(H)} 
                    & \footnotesize{Q(He)}
                    & \footnotesize{Q(He$^{+}$)} \\
\tableline
\footnotesize  500   &   \footnotesize 6.802e50   &   \footnotesize 3.858e50   &   \footnotesize 5.793e49   \\
\footnotesize  400   &   \footnotesize 5.247e50   &   \footnotesize 3.260e50   &   \footnotesize 5.567e49   \\
\footnotesize  300   &   \footnotesize 3.754e50   &   \footnotesize 2.372e50   &   \footnotesize 4.190e49   \\
\footnotesize  200   &   \footnotesize 2.624e50   &   \footnotesize 1.628e50   &   \footnotesize 1.487e49   \\
\footnotesize  120   &   \footnotesize 1.391e50   &   \footnotesize 7.772e50   &   \footnotesize 5.009e48   \\
\tableline
\end{tabular}
\end{center}
\vspace{0.1in}
\end{tablehere}

\subsection{Halo Mass}

An interesting question is whether a sufficiently massive halo is capable
of trapping the HII region. The escape velocity has no direct effect on 
escape fraction because the fraction depends only on the reversion
of the front back to R-type, not on whether any gas exits the halo. This 
transition will occur in any halo whose baryonic density
drops off more steeply than r$^{-1.5}$, whether or not there is outflow 
from the halo (the Abel, et al densities as well as more recent $\Lambda$CDM 
profiles all decrease approximately as r$^{-2.5}$).

However, core densities could be sufficiently high in massive halos to 
prevent this transition from occurring before the main sequence lifetime 
of the central source, in which case the front would be trapped as an 
ultracompact (UC) HII region \citep{kurtz00}. 
We tested three halo mass profiles with a 200 M$_{\odot}$
star by scaling the original density profile of \citet{abn02} by factors 
of 2 and 5 (n(r/2) and n(r/5)). The most massive halo confined the I--front
to a 0.1 pc UC HII region while the n(r/2) halo permitted the front to
escape, as seen clearly in Figs 7 and 8. The front has slightly higher 
temperatures in the greater n(r/2) density and thus develops higher shock
speeds.  However, the delay in the D- to R-type transition keeps the core 
shock from reaching the same radius as in the original distribution.

The high central densities of the most massive halo prevent the reversion 
of the I-front from D-type to R-type before the star exits 
the main sequence.  Only in this unrealistically limiting case does no UV 
(or therefore gas) escape the halo.  Breakout occurs in the other two cases 
although it occurs later in the more massive of the two halos so less mass 
is ejected from it.  The final I-sphere radius of the n(r/2) halo is also 
smaller than the n(r) I-sphere because the n(r/2) halo exhibits its high 
f$_{esc}$ for less time. Until the relationship between the mass of a 
minihalo and the central star that may accrete within it is better 
understood it will not be possible to make a precise determination of how 
many Pop III I-fronts achieve breakout, only that the majority of them do 
for a reasonable scaling of halo densities with stellar mass.

\section{Conclusions}

Realistic escape fractions from the first luminous objects are close
to unity because the halos become nearly transparent to UV photons
when ionized by the R-type front. This result is of particular
importance because recent determinations of the electron-scattering
optical depth of the cosmic microwave background radiation since
recombination suggest an early period of reionization \citep{ket03}.
The very massive stars (30 M$_\odot$ $\lesssim$ M$_{FS}$ $\lesssim$ 300
M$_\odot$) emerging in recent numerical simulations in concert with
large UV escape fractions may likely be large contributors to early
reionization. The photoevaporative flows of the first stars will leave
dark halos with low gas content regardless of whether these stars die
in bright supernovae or form black holes (e.g. \citet{hw02}, for an
extended discussion). Since the typical expansion velocities of the
flows we find are much larger than the halo escape velocity one cannot
expect the gas to return within timescales in which the halo would
merge again into larger objects. Black hole remnants of the first
stars that created the HII region could not accrete significant mass
in the low density environment their progenitors have created. This
fact may be an important constraint for theories of supermassive black
hole formation relying on constant feeding of stellar black hole
remnants (e.g. \citet{hl01} and references therein).

The HII region that evacuates the halo will also enhance metal
enrichment of the halo and interhalo medium by the first supernova. A 200 M$_\odot$
pair-instability supernova remnant initially expands into an ambient
density that is only twice the interhalo mean. Fig 5 shows this
ejecta must travel 10 pc before encountering its own mass in halo
material, before which it would be in a free-expansion state. The
remnant will not enter a Sedov-Taylor phase until it is at least 50 pc
in radius, at which point Rayleigh-Taylor instabilities that we cannot
simulate in 1D begin to mix the ejected metals with the gas in the
halo. How these instabilities promote metal pollution of the early IGM
will be the focus of future study. Only three dimensional
models of the first HII regions will allow us to address the detailed questions
recently raised by \citet{oh03}.

\acknowledgments
We would like to thank the anonymous referee for suggestions that significantly
improved the quality of this manuscript.  This work was partially supported by NSF 
grants AST-9803137 and AST-0307690 to MLN. The calculations were carried out on NCSA's 
Origin2000 with NPAC allocation PQK.

\end{document}